\documentclass[prb,aps,twocolumn,floatfix,tightenlines,amsmath,amssymb,showpacs,superscriptaddress,notitlepage,longbibliography]{revtex4-2}
\usepackage[T1]{fontenc}
\usepackage{lmodern}
\usepackage{amsmath,amssymb,mathtools,bm}
\usepackage{graphicx}
\usepackage{microtype}
\usepackage{xcolor}
\usepackage[hidelinks]{hyperref}
\graphicspath{{figures/}}

\newcommand{\up}{\uparrow}
\newcommand{\dn}{\downarrow}

\begin{document}

\title{Theory of spin-polarized scanning tunneling microscopy with topological insulator tips}

\author{Zhuo-Yuan Shi}
\affiliation{Center for Correlated Matter and School of Physics, Zhejiang University, Hangzhou 310058, China}

\author{Chuang Li}
\email{lichuang.zju@outlook.com}
\affiliation{Center for Correlated Matter and School of Physics, Zhejiang University, Hangzhou 310058, China}

\author{Lin Jiao}
\affiliation{Center for Correlated Matter and School of Physics, Zhejiang University, Hangzhou 310058, China}

\author{Huiqiu Yuan}
\affiliation{Center for Correlated Matter and School of Physics, Zhejiang University, Hangzhou 310058, China}

\author{Zhenyu Wang}
\affiliation{Department of Physics, University of Science and Technology of China, Hefei, Anhui, China}
\affiliation{Hefei National Laboratory, University of Science and Technology of China, Hefei, 230088, China}

\author{Xianhui Chen}
\affiliation{Department of Physics, University of Science and Technology of China, Hefei, Anhui, China}
\affiliation{Hefei National Laboratory, University of Science and Technology of China, Hefei, 230088, China}

\author{Lun-Hui Hu}
\email{lunhui@zju.edu.cn}
\affiliation{Center for Correlated Matter and School of Physics, Zhejiang University, Hangzhou 310058, China}

\date{\today}

\begin{abstract}
Topological insulators have enabled a new class of spin-sensitive scanning tunneling microscopy (STM) probes based on helical boundary states~\cite{Aishwarya2022,Wang2025HelicalTip,Yang2026AltermagnetTip}. However, how these probes generate a net spin-polarized tunneling current remains unclear. We present a theory of spin-polarized STM with topological insulator tips and show that asymmetric tunneling-contact geometry and spin dephasing enable helical boundary states to generate a spin-polarized tunneling current. Within the standard Landauer-B\"uttiker formalism, we demonstrate this mechanism for a quantum spin Hall (QSH) insulator tip: asymmetric contacts make the two counter-propagating helical edge states, which carry opposite spins, traverse path lengths that differ by $\Delta L$, and spin dephasing induces backscattering that reduces the conductance more strongly for the longer path. For a representative asymmetric contact and weak spin dephasing, the QSH tip exhibits a tunneling spin polarization of $P_s=-17.2\%$ under positive bias, whose magnitude can reach $50\%$ for larger $\Delta L$, and which remains robust against Rashba spin-orbit coupling. Crucially, this polarization, dominated by helical edge transport rather than by bulk states, naturally reverses sign with the bias polarity owing to spin-momentum locking. Our findings are consistent with recent STM experiments using topological tips.
\end{abstract}

\maketitle

\section{Introduction}

Topological quantum materials have opened new opportunities for quantum transport and device applications owing to their nontrivial band topology and low-dissipation transport properties~\cite{HasanKane2010,QiZhang2011,Chiu2016RMP,Bansil2016RMP,Armitage2018RMP,lv2021RMP,Wieder2022NRM,Chang2023RMP}. Recent high-throughput computations have identified thousands of topological materials, which are now catalogued in open databases~\cite{Zhang2019Nature,Tang2019Nature,Vergniory2019Nature}. However, realizing direct functional applications typically requires the boundary-state transport to dominate at the Fermi level, though a fully gapped bulk is not strictly necessary. Among this broad family, topological insulators naturally satisfy this criterion, combining an insulating bulk with gapless boundary states that are protected by time-reversal symmetry~\cite{KaneMele2005,BernevigHughesZhang2006,Konig2007,FuKaneMele2007,Hsieh2008,Zhang2009,Chen2009,Xia2009}. Spin–momentum locking converts a charge current in these topological boundary states into a net spin polarization, and the robustness of these states against nonmagnetic disorder makes this spin-sensitive transport highly attractive for promising low-dissipation spintronic devices~\cite{Pesin2012NM,Han2018npj,He2022NM,Breunig2022NRP,Que2026AM}. In addition to their considerable device potential, helical boundary states of topological insulators have recently afforded an unexpected and distinct application: spin-sensitive probes of quantum matter~\cite{Aishwarya2022,Wang2025HelicalTip,Yang2026AltermagnetTip}.

A striking realization of this probe concept is scanning tunneling microscopy (STM) using topological-insulator tips~\cite{Aishwarya2022,Wang2025HelicalTip,Yang2026AltermagnetTip}. In this approach, spin-selective tunneling originates from spin-momentum-locked helical boundary states within the tip itself. The first demonstration employed SmB$_6$ nanowires, a candidate topological Kondo insulator~\cite{Dzero2010PRL,Xu2014SpinTexture,Dzero2016,Jiao2018MagneticDefect,Li2020NRP}, as STM tips to image the bicollinear antiferromagnetic order of Fe$_{1.03}$Te with atomic resolution at 4 K~\cite{Aishwarya2022}. The magnetic contrast vanished upon warming above $\sim 10$ K, concomitant with the suppression of the topological surface states in the SmB$_6$ tip~\cite{Aishwarya2022,Neupane2013NC}, well below the N\'eel temperature ($\sim 50$ K) of Fe$_{1.03}$Te. Crucially, reversing the STM bias polarity produced a $\pi$ phase reversal of the magnetic stripes, showing that the spin selectivity of the tip can be reversed purely electrically.

More recently, another topological insulator, Bi$_4$Br$_4$~\cite{Zhou2014NL,Noguchi2021,Shumiya2022,Lin2024NC,Yu2024PRX}, with helical boundary states, has also been used as an STM tip to image spin-resolved quasiparticle interference (QPI) in the $d$-wave layered altermagnet KV$_2$Se$_2$O~\cite{Wang2025HelicalTip}. Strikingly, reversing the STM bias rotates the dominant QPI pattern by $90^\circ$. In a $d$-wave altermagnet, the spin splitting changes sign under such a rotation because of the $[C_2||C_4]$ symmetry~\cite{Smejkal2022PRXLandscape,Bai2024AFM,Fender2025JACS,Jungwirth2026Nature}; the observed rotation thus directly indicates that the spin selectivity of the Bi$_4$Br$_4$ tip reverses sign with bias polarity. The same bias-driven reversal of spin-selective QPI on KV$_2$Se$_2$O has also been independently observed with SmB$_6$ tips~\cite{Yang2026AltermagnetTip}. Meanwhile, conventional spin-polarized STM using ferromagnetic Cr tips has observed QPI patterns with the same $90^\circ$ rotation on KV$_2$Se$_2$O and its sister compound RbV$_2$Se$_2$O, albeit with the contrast switched by an applied magnetic field rather than by the bias~\cite{Gu2026arXiv,Mu2026arXiv}. This underscores the distinctive capability of topological tips to achieve comparable spin selectivity through bias reversal alone, without tip magnetization. Nevertheless, the microscopic mechanism by which helical boundary states generate a spin-polarized tunneling current remains unclear.

\begin{figure}[t]
\centering
\includegraphics[width=0.48\textwidth]{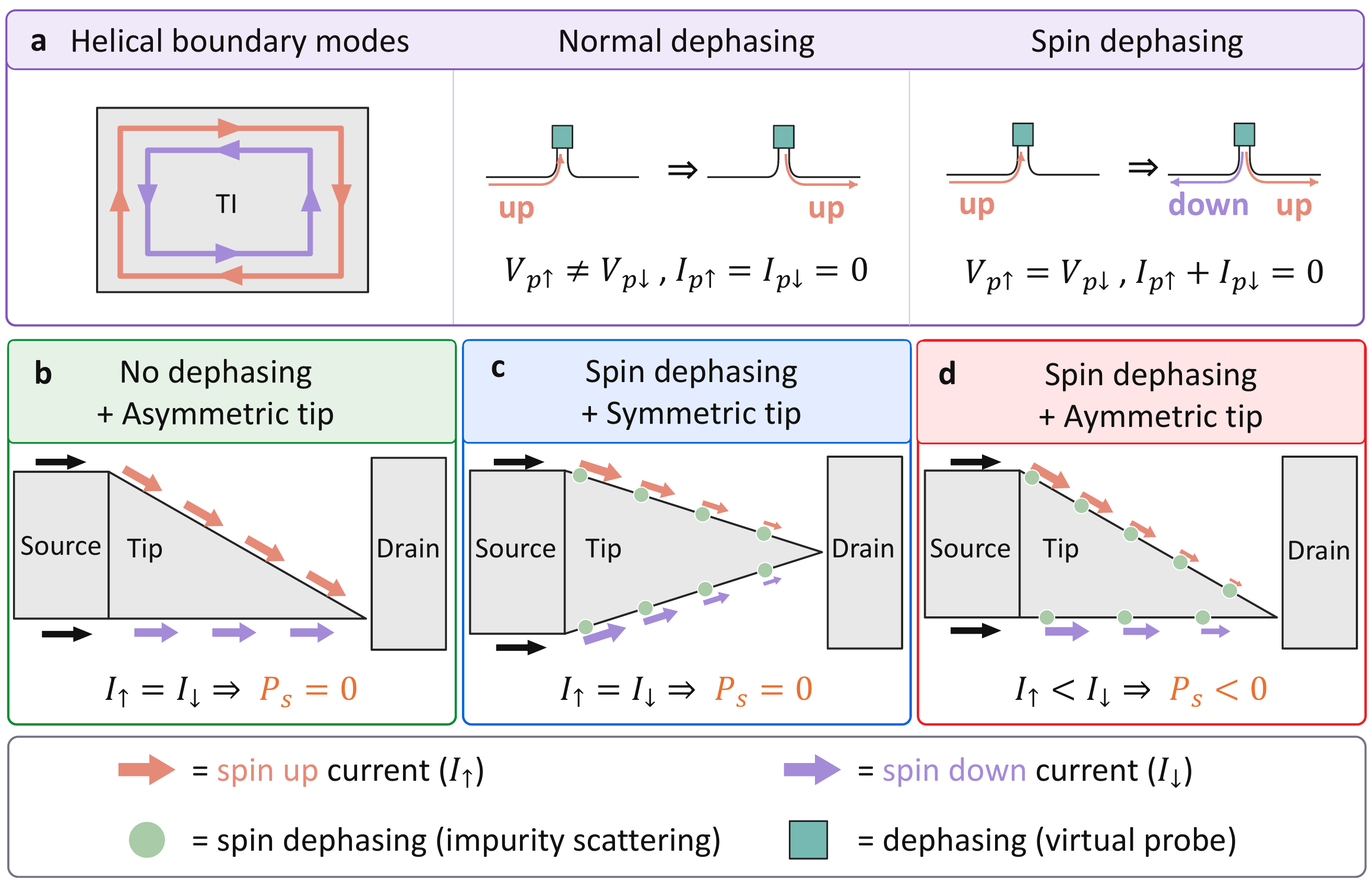}
\caption{Mechanism of spin-selective tunneling in topological-insulator tips. 
(a) Comparison of normal and spin dephasing in helical boundary modes with opposite spins. Normal (spin-conserving) dephasing produces no backscattering of the edge states, whereas spin-relaxing dephasing induces spin-flip backscattering.
(b)--(d) Three main cases illustrating the requirements for tunneling spin polarization: 
(b) asymmetric geometry without dephasing (or with normal dephasing only), yielding unequal path lengths but balanced transmissions; 
(c) symmetric geometry with spin dephasing, where both channels are equally suppressed; 
and (d) the combination of asymmetric geometry and spin dephasing, where unequal spin-flip scattering generates a finite spin-polarized current.
}
\label{fig:mechanism}
\end{figure}

Since a topological-insulator tip preserves time-reversal symmetry, resolving this puzzle requires a theoretical framework distinct from that of conventional ferromagnetic tips, whose spin selectivity mainly originates from unequal spin-resolved densities of states~\cite{TersoffHamann1983,Wiesendanger1990,Bode2003,Wiesendanger2009,Oka2014RMP}. Specifically, in such a tip, the two counter-propagating helical modes of opposite spin give exactly balanced spin-resolved densities of states [Fig.~\ref{fig:mechanism}(a)]. The ferromagnetic-tip framework therefore predicts vanishing spin selectivity and cannot explain the observed bias-reversal effect. Importantly, because an STM tip terminates in a sharp apex, the two helical boundary states couple to the sample at different distances, making the tunneling-contact geometry asymmetric [see Fig.~\ref{fig:mechanism}]. To model this transport, we employ a nonequilibrium Landauer-B\"uttiker formalism with current-conserving B\"uttiker probes~\cite{Buttiker1986,DAmatoPastawski1990,Jiang2009,GuoSun2012,Erhardt2025NC} that treat spin dephasing as spin-relaxing backscattering [Fig.~\ref{fig:mechanism}(a)].

Our central finding is that a net spin-polarized tunneling current from topological-insulator tips arises only when the asymmetric contact geometry and spin dephasing coexist and act together. First, the tip-apex-induced asymmetric tunneling-contact geometry (without spin dephasing) creates helical propagation paths whose lengths differ by $\Delta L$, but does not induce backscattering of the edge modes, leaving the spin-resolved transmissions balanced [Fig.~\ref{fig:mechanism}(b)]. Second, with spin dephasing but symmetric propagation paths, both helical channels experience identical suppression, so the spin-resolved transmissions remain balanced [Fig.~\ref{fig:mechanism}(c)]. Only when the two act together do the helical channels acquire unequal transmission probabilities and thus a net tunneling spin polarization ($P_s \neq 0$) [Fig.~\ref{fig:mechanism}(d)]. Moreover, the finite $P_s$ is dominated by helical edge transport, reverses sign with the bias polarity owing to spin-momentum locking, and can be substantially enhanced by increasing $\Delta L$. As a concrete demonstration, we employ a quantum spin Hall (QSH) insulator tip to quantify the resulting spin polarization, obtaining $P_s \approx -17.2\%$; these results are qualitatively consistent with recent experiments on topological STM probes~\cite{Aishwarya2022,Wang2025HelicalTip,Yang2026AltermagnetTip} and provide design principles for optimizing the spin sensitivity of such probes.

\section{Spin-resolved Landauer-B\"uttiker formalism}
\label{sec:formalism}

The standard theory of STM is built on Bardeen's transfer-Hamiltonian formalism, which treats electron tunneling perturbatively to leading order in the tip-sample coupling~\cite{Bardeen1961,Chen1993STM,Wiesendanger1994SPM,Voigtlaender2015SPM}. In its application to spin-polarized STM with ferromagnetic tips, the tunneling matrix element is taken to be spin independent, and the spin-resolved current is given by $I_\sigma(U) \propto \int dE\, |M|^2\, \rho_{t,\sigma}(E - eU)\, \rho_{s,\sigma}(E)\, \big[f(E - eU) - f(E)\big]$, where $\sigma = \uparrow,\downarrow$ labels the spin of the tunneling electron, $U$ is the bias voltage applied between the tip and the sample, $M$ is the tunneling matrix element, $\rho_{t,\sigma}(E)$ and $\rho_{s,\sigma}(E)$ are the spin-resolved densities of states of the tip and the sample, respectively, and $f(E) = [e^{E/(k_{\mathrm{B}}T)} + 1]^{-1}$ is the Fermi distribution function. Within this description, the spin polarization of the tunneling current, $P_s = (I_\uparrow - I_\downarrow)/(I_\uparrow + I_\downarrow)$, originates entirely from the imbalance between $\rho_{t,\uparrow}$ and $\rho_{t,\downarrow}$~\cite{TersoffHamann1983,Wiesendanger1990,Bode2003,Wiesendanger2009,Oka2014RMP}. This perturbative picture, however, is inadequate for a topological-insulator tip~\footnote{One might instead attribute the spin selectivity to a spin-dependent matrix element computed from the helical wavefunctions of the tip. Such an elastic matrix element, however, is independent of the bias voltage and cannot account for the experimentally observed sign reversal of the spin signal when the bias is reversed.}: the spin signal is carried by helical boundary states that propagate along the tip before tunneling, and their transmission is shaped by repeated scattering within the tip and by spin dephasing along the edge paths [see Fig.~\ref{fig:mechanism}], effects that a leading-order tunneling matrix element cannot capture~\cite{Feuchtwang1974,SacksNoguera1991,Wortmann2005}. We therefore employ the Landauer-B\"uttiker formalism within the nonequilibrium Green-function approach, which describes transport as phase-coherent transmission between reservoirs and incorporates dephasing through current-conserving B\"uttiker probes~\cite{Buttiker1986,DAmatoPastawski1990,Jiang2009,GuoSun2012,Erhardt2025NC}.

As illustrated in Fig.~\ref{fig:mechanism}(d), we consider transport from the topological-insulator tip (the left lead $L$), through the central scattering region, to the sample (the right lead $R$). To implement dephasing, we attach B\"uttiker probes to $N$ selected lattice sites of the topological-insulator tip. At each selected site $p$, the probe modifies the tip Hamiltonian by adding a spin-resolved self-energy, 
\begin{align}
\Sigma_{p\sigma}^r=-i\beta\Pi_{p\sigma}/2,    
\end{align}
where $\Pi_{p\sigma}$ projects onto all orbitals of spin $\sigma$ at site $p$ (i.e.,~the spin-$\sigma$ block of the on-site Hilbert space), and $\beta\geq0$ is the probe strength (i.e.,~dephasing strength). We consider two types of dephasing mechanisms~\cite{Jiang2009}: (i) normal dephasing (spin-conserving) imposes $I_{p\up}=I_{p\dn}=0$ so that generally $V_{p\up}\neq V_{p\dn}$, and (ii) spin dephasing (spin-relaxing) imposes $V_{p\up}=V_{p\dn}$ and $I_{p\up}+I_{p\dn}=0$. Here $V_{p\sigma}$ and $I_{p\sigma}$ denote the voltage and current of the corresponding B\"uttiker probe.

To determine the spin selectivity of the topological-insulator tip, we model the sample as a spin-degenerate metal without spin-orbit coupling, so that any spin polarization of the current can be attributed solely to the tip. All terminals, including the B\"uttiker probes, are then treated on an equal footing within the multiterminal Landauer-B\"uttiker formalism. The transmission from terminal $j$ to terminal $i$ is given by the Caroli formula~\cite{Caroli1971},
\begin{equation} \label{eq:green_trans}
T_{i\leftarrow j}(E)=\operatorname{Tr}[\Gamma_iG^r(E)\Gamma_jG^a(E)],
\end{equation}
where $E$ is the electron energy, $G^{r/a}(E)$ are the retarded/advanced Green functions of the central region dressed by the self-energies of the physical leads ($L$, $R$) and the local B\"uttiker probes ($p=1,2,\cdots,N$), and $\Gamma_i=i(\Sigma_i^r-\Sigma_i^a)$ is the broadening matrix~\cite{Datta2005QuantumTransport}. Here $i=(\alpha,\sigma)$ and $j=(\alpha',\sigma')$ are composite terminal-spin indices, with $\alpha,\alpha'\in\{L,R,1,2,\ldots,N\}$ and $\sigma,\sigma'=\{\uparrow,\downarrow\}$. The trace runs over all central-region spatial and orbital degrees of freedom. As a result, the probe voltages $V_{p\sigma}$ are determined self-consistently from the spin-resolved multiterminal Landauer-B\"uttiker equations~\cite{GuoSun2012},
\begin{align} \label{eq:current-itoj}
I_i= \frac{e^2}{h} \sum_{j\ne i}(T_{i\leftarrow j}V_j-T_{j\leftarrow i}V_i),
\end{align}
where $I_i$ is the current entering terminal $i$: the first term collects electrons injected from $j$ into $i$, and the second collects electrons leaving $i$ towards $j$. At zero temperature we set $V_{L\sigma}=V$ and $V_{R\sigma}=0$, with $V$ the applied STM bias. Within linear response, the current collected by the right lead is $I_{R\sigma}(E)=(e^2/h)T^{\mathrm{eff}}_{R\sigma\leftarrow L}(E)V$, with the effective spin-resolved transmission
\begin{equation}  \label{eq:eff_trans}
T^{\mathrm{eff}}_{R\sigma\leftarrow L}(E)=\sum_{\sigma'}\left[T_{R\sigma\leftarrow L\sigma'}+\sum_p T_{R\sigma\leftarrow p\sigma'}\frac{V_{p\sigma'}}{V}\right].
\end{equation}
Eq.~\eqref{eq:eff_trans} is the central quantity of our theory. Its first term, the Bardeen-type term, collects all direct tunneling processes from the tip to the sample. Its second term, the B\"uttiker-probe-mediated term, collects the indirect processes in which an electron is absorbed by a B\"uttiker probe, reinjected at the probe voltage $V_{p\sigma'}$ with or without spin flip according to the dephasing condition, and subsequently transmitted to the sample. The spin polarization of the tunneling current~\cite{Sun2026MolecularKnots} is then given by
\begin{equation}  \label{eq:spin_polarization}
P_s(E)=\frac{T^{\mathrm{eff}}_{R\up\leftarrow L}(E)-T^{\mathrm{eff}}_{R\dn\leftarrow L}(E)}{T^{\mathrm{eff}}_{R\up\leftarrow L}(E)+T^{\mathrm{eff}}_{R\dn\leftarrow L}(E)}.
\end{equation}

Reversing the bias polarity amounts to interchanging $L\leftrightarrow R$ in Eq.~\eqref{eq:eff_trans}. While $P_s(E)$ defined in Eq.~\eqref{eq:spin_polarization} characterizes the injection from the tip to the sample ($L \to R$), reversing the bias corresponds to the reverse injection process ($R \to L$), which we denote as $\bar{P}_s(E)$. Owing to the spin-momentum locking of the helical modes, reversing the transport direction reverses the spin orientation carried by the edge modes. Consequently, we expect the spin polarization to reverse its sign upon reversing the bias polarity, i.e.,
\begin{align} \label{eq:spinPs-reversal}
\operatorname{sign}\!\left[P_s(E)\bar{P}_s(E)\right]<0.
\end{align}
This distinctive prediction is confirmed by the calculations presented in Sec.~\ref{sec:direction}, and further details are provided in the Supplemental Material (SM)~\cite{SM}.

Before model calculations, we make a qualitative comparison between ferromagnetic and topological-insulator tips. In a ferromagnetic tip, the B\"uttiker probes are absent, so Eq.~\eqref{eq:eff_trans} reduces to $T^{\mathrm{eff}}_{R\sigma\leftarrow L}=T_{R\sigma\leftarrow L} \propto \rho_{t,\sigma}$, where $\rho_{t,\sigma}$ is the spin-resolved density of states of the tip. In a topological-insulator tip, the B\"uttiker probe-mediated second term in Eq.~\eqref{eq:eff_trans} drives the spin selectivity: spin dephasing redistributes the two spin populations, which, combined with the asymmetric edge paths, yields unequal effective transmissions and a finite polarization.

\begin{figure*}[t]
\centering
\includegraphics[width=0.98\textwidth]{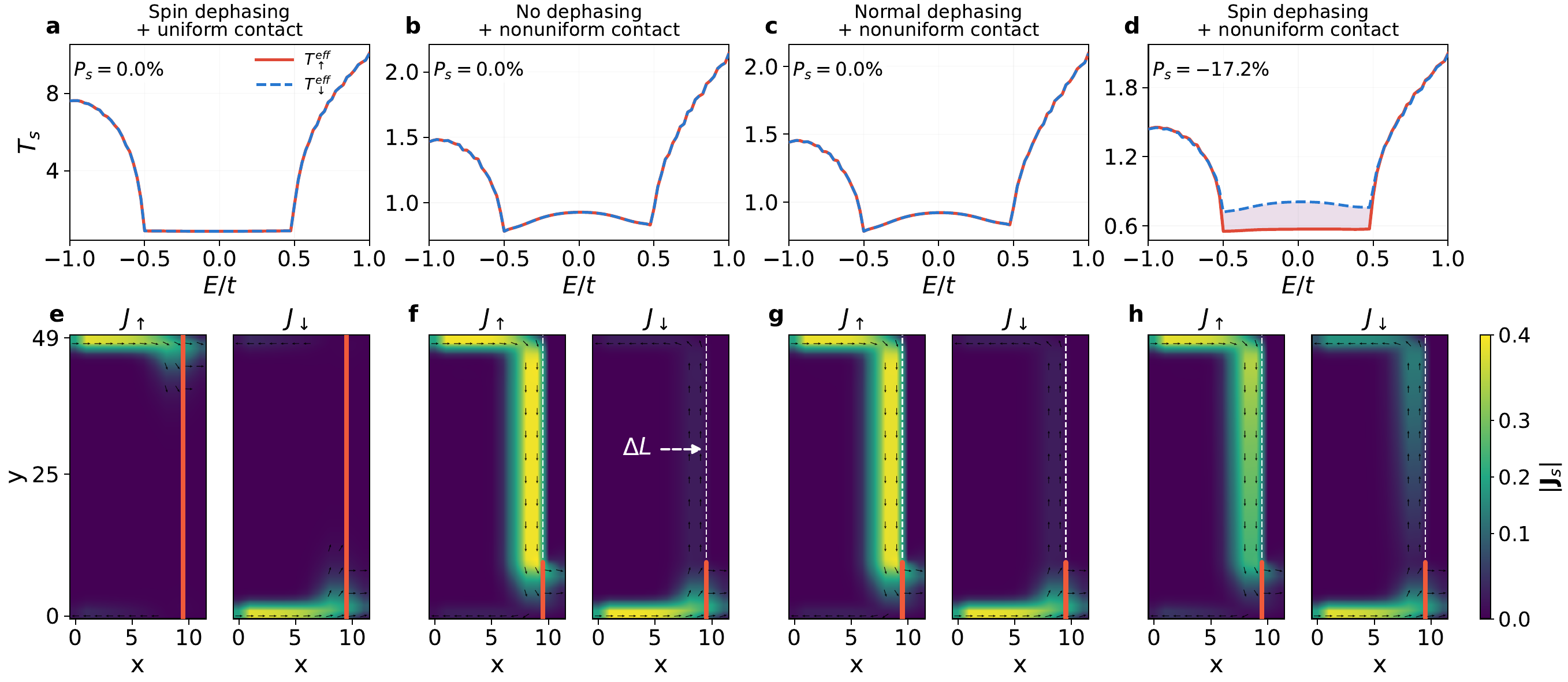}
\caption{Spin-resolved transport and local currents for different contact geometries and dephasing mechanisms. 
(a)--(d) Spin-resolved effective transmission spectra $T^{\mathrm{eff}}_{R\uparrow\leftarrow L}(E)$ (solid red) and $T^{\mathrm{eff}}_{R\downarrow\leftarrow L}(E)$ (dashed blue) for four configurations:
(a) symmetric contact with spin dephasing, 
(b) asymmetric contact without dephasing, 
(c) asymmetric contact with normal dephasing, 
and (d) asymmetric contact with spin dephasing. 
Only the coexistence of an asymmetric geometry and spin dephasing (d) yields unequal spin-resolved transmissions within the bulk gap ($|E|<0.5t$). 
(e)--(h) Real-space distributions of the spin-resolved local currents $J_\uparrow$ (left) and $J_\downarrow$ (right) at $E=0.01t$, corresponding to the four conditions in (a)--(d), respectively. At the tip-sample interface, solid orange lines denote active tunnel contacts, while dashed white lines denote regions without tunneling. Colors encode the magnitude $|J_s|$, and arrows (normalized for visibility) indicate the local current direction. An enhanced spin-flip return flow along the longer upper edge is visible in (h).
}
\label{fig:controls}
\end{figure*}

\section{Model and numerical results}
\label{sec:controls}

The general formalism of Sec.~\ref{sec:formalism} identifies the conditions under which a topological-insulator tip can generate a spin-polarized current, but it cannot by itself answer three quantitative questions: how large the polarization is, how it depends on the contact geometry and the dephasing strength, and whether it is indeed carried by helical edge transport within the bulk gap rather than by bulk states. To address these questions, we construct in this section a minimal lattice model of a topological-insulator tip with an asymmetric tunneling contact and a controlled placement of the B\"uttiker probes, and compute the spin polarization $P_s$ as a function of the edge-path difference $\Delta L$ and the dephasing strength $\beta$. As the minimal model hosting counter-propagating helical boundary states, we employ a quantum spin Hall (QSH) insulator tip~\cite{Wang2025HelicalTip}.

\subsection{QSH tip Hamiltonian and contact geometry}

For concreteness, we model the QSH tip and the normal-metal sample on a square lattice with lattice constant $a=1$. At each site $\bm r=(x,y)$, $c_{\bm r}$ is the four-component electron annihilation operator in the basis $\{ |s,\up\rangle, |p_-,\up\rangle, |s,\dn\rangle, |p_+,\dn\rangle \}$. The Pauli matrices $\sigma_i$ and $\tau_i$ ($i=x,y,z$) act on spin and orbital spaces, respectively, while $\sigma_0$ and $\tau_0$ denote the identity matrices. The real-space BHZ Hamiltonian reads~\cite{BernevigHughesZhang2006},
\begin{equation}  \label{eq:bhz}
H_{\rm QSH}
=\sum_{\bm r}c_{\bm r}^{\dagger} h_0 c_{\bm r}
+\sum_{\bm r,\hat{\mu}=x,y}
\left(c_{\bm r+\hat{\mu}}^{\dagger}T_{\hat{\mu}} c_{\bm r}
+\mathrm{H.c.}\right),
\end{equation}
where $h_0=(M-4t)\sigma_0\tau_z$ is the on-site term, and $T_x=t \sigma_0\tau_z-\frac{iA}{2}\sigma_z\tau_x$ and $T_y=t \sigma_0\tau_z-\frac{iA}{2}\sigma_0\tau_y$ are the nearest-neighbor hopping matrices along the $x$ and $y$ directions, respectively. Here, $M$ controls the band inversion, $t$ is the orbital-dependent hopping amplitude, and $A$ sets the interorbital hopping strength. The QSH phase emerges when the mass term is inverted at the $\Gamma$ point but not at the $M$ point, which requires $0 < M < 4t$ (assuming $t>0$). Moreover, the Hamiltonian preserves time-reversal symmetry and conserves $\sigma_z$ (i.e., $[H_{\mathrm{QSH}},\sigma_z\tau_0]=0$), so the two spin blocks decouple and the spin-resolved analysis below is unambiguous. Within the bulk gap $|E| < |M|$, the system hosts a single Kramers pair of helical edge states [see the SM~\cite{SM}], which mediate the spin-selective transport studied in this work. The effect of Rashba spin-orbit coupling, which breaks the $\sigma_z$ conservation, is discussed in Sec.~\ref{sec:rashba}.

The normal-metal sample contains two degenerate orbitals, with the electron annihilation operator $d_{\bm r}=(d_{\bm r,1,\up},d_{\bm r,2,\up},d_{\bm r,1,\dn},d_{\bm r,2,\dn})^{\mathsf T}$. The Hamiltonian reads
\begin{equation}
H_{\rm NM}
=\epsilon_m \sum_{\bm r}d_{\bm r}^{\dagger}d_{\bm r}
-t_m\sum_{\bm r,\hat{\mu}=x,y}
\left(d_{\bm r+\hat{\mu}}^{\dagger}d_{\bm r}
+\mathrm{H.c.}\right),
\label{eq:normal_metal}
\end{equation}
where $\epsilon_m=4t_m-\mu_m$ is the on-site energy, $\mu_m$ is the chemical potential, and $t_m$ is the nearest-neighbor hopping amplitude. The system is semi-infinite along the $x$ direction and has a width of $N_y$ sites along the $y$ direction. For the transport calculation, we retain a finite central region comprising a tip segment ($0 \leq x \leq L_{\rm QSH}$) and a sample segment ($L_{\rm QSH}+1 \leq x \leq L_{\rm QSH}+L_{\rm NM}$), while the remaining semi-infinite portions are incorporated through the self-energies in Eq.~\eqref{eq:green_trans}. The tip-sample coupling at the interface $x=L_{\rm QSH}$ is given by
\begin{align} \label{eq:contact}
H_{\rm contact}
=-t_c\sum_{y\in\mathcal W_c}
\left(d_{L_{\rm QSH}+1,y}^{\dagger}c_{L_{\rm QSH},y}
+\mathrm{H.c.}\right),
\end{align}
where $t_c$ is the coupling constant, and the set $\mathcal W_c$ specifies the contacted sites. Specifically, (i) a symmetric (uniform) contact corresponds to $\mathcal W_c=\{0,\ldots,N_y-1\}$, whereas (ii) an asymmetric (nonuniform) contact is defined by $\mathcal W_c=\{0,\ldots,N_c-1\}$ with $1 \leq N_c \leq N_y-1$. The asymmetric contact captures the finite size and sharp apex of an STM tip, so that the two helical edge states travel different effective distances before reaching the sample. This path asymmetry is quantified by
\begin{align}
\Delta L=N_y-N_c.
\end{align}

To further incorporate dephasing along the helical edge propagation paths, we attach local B\"uttiker probes to the outermost three lattice layers along the upper, lower, and right boundaries of the QSH segment within the central region, with the dephasing strength quantified by the parameter $\beta$. Finally, we apply Eq.~\eqref{eq:eff_trans} and Eq.~\eqref{eq:spin_polarization} to compute the effective transmissions and the resulting spin polarization, respectively. The model parameters are set to $M=0.5$, $t=1$, $A=1$, $t_m=1$, $\mu_m=2$, and $t_c=0.6$.

\subsection{Spin-resolved transmission spectra}

Figure~\ref{fig:controls}(a)–(d) display the calculated spin-resolved effective transmission spectra $T^{\mathrm{eff}}_{R\uparrow\leftarrow L}(E)$ and $T^{\mathrm{eff}}_{R\downarrow\leftarrow L}(E)$ for four distinct configurations. Inside the bulk gap ($|E|<0.5$), transport is governed by the helical edge modes. A key observation is that neither the contact asymmetry nor the dephasing mechanism alone yields appreciable spin selectivity. For a symmetric contact ($\Delta L = 0$), spin dephasing suppresses both helical channels identically, preserving a vanishing polarization [Fig.~\ref{fig:controls}(a)]. Similarly, when $\Delta L \neq 0$, an asymmetric contact either without dephasing [Fig.~\ref{fig:controls}(b)] or with normal dephasing [Fig.~\ref{fig:controls}(c)] leaves the two spin-resolved transmissions strictly degenerate across the bulk gap. 

In stark contrast, the interplay between contact asymmetry and spin-relaxing dephasing lifts this degeneracy [Fig.~\ref{fig:controls}(d)]. For $\Delta L = 40$ and $\beta = 0.03$, the effective transmission for spin-up is significantly reduced relative to spin-down, yielding a net spin polarization of $P_s = -17.2\%$ at $E=0.01$. The negative sign of $P_s$ is a direct consequence of the specific chirality (spin-momentum locking) of the helical edge modes in our model. The spin-up and spin-down electrons form clockwise and counterclockwise propagating modes along the boundaries. Since the asymmetric contact makes the shorter lower edge more transmissive, the net spin polarization is dominated by the spin-down electrons propagating along this edge towards the contact, yielding $P_s<0$.

To explicitly verify this microscopic origin, we reverse the chirality of the edge states by flipping the sign of $A$ solely in the $y$-direction hopping $T_y$ (i.e., $A \to -A$ while keeping $T_x$ unchanged) in Eq.~\eqref{eq:bhz}. This operation swaps the propagation directions of the spin-up and spin-down modes. Under the identical asymmetric contact geometry and spin-relaxing dephasing, this chirality reversal exactly flips the dominant spin species, resulting in a positive spin polarization of $P_s = +17.2\%$ [see the SM~\cite{SM}]. Therefore, these results establish that both contact asymmetry and spin relaxation are essential prerequisites for spin-selective tunneling from a QSH tip, with the sign of $P_s$ serving as a direct fingerprint of its spin-momentum locking orientation.

\subsection{Spatial distributions of spin-resolved currents}

To further elucidate the real-space mechanism underlying this spin-selective transport, we examine the spatial distributions of the spin-resolved local currents at $E=0.01$, as shown in Figs.~\ref{fig:controls}(e)–(h). For a symmetric contact ($\Delta L=0$) with spin dephasing, both helical edge pathways are geometrically equivalent, so their spin-relaxing attenuations remain strictly balanced, yielding equal spin-up and spin-down currents [Fig.~\ref{fig:controls}(e)]. For an asymmetric contact ($\Delta L \neq 0$) in the absence of spin dephasing or with normal dephasing, elastic propagation along the extended upper edge is lossless, and spin remains locked to momentum all the way to the sample. The two helical channels therefore contribute equal spin-resolved currents, preserving a vanishing net polarization [Figs.~\ref{fig:controls}(f) and \ref{fig:controls}(g)]. A pronounced asymmetry emerges exclusively when spin-relaxing dephasing coexists with the asymmetric contact [Fig.~\ref{fig:controls}(h)]. Along the geometrically extended upper edge, electrons undergo a longer propagation distance, thereby accumulating stronger spin-flip backscattering. This unequal spin-flip scattering leads to unequal spin-resolved transmissions reaching the sample, consistent with the calculated spin polarization.

Further details, including the local current calculations, the parameter dependence of $P_s$ ($\Delta L$ and $\beta$), and its weak sensitivity to the contact hopping $t_c$ [see Eq.~\eqref{eq:contact}], are provided in the SM~\cite{SM}.

\begin{figure}[t]
\centering
\includegraphics[width=0.48\textwidth]{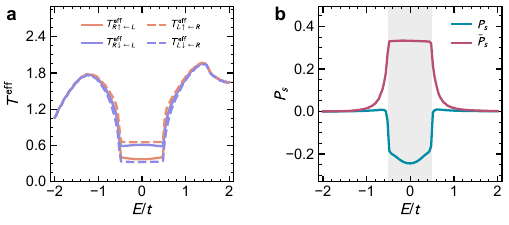}
\caption{Bias-polarity-induced reversal of tunneling spin polarization.
(a) Spin-resolved effective transmissions $T^{\mathrm{eff}}_{R\uparrow\leftarrow L}(E)$ (orange) and $T^{\mathrm{eff}}_{R\downarrow\leftarrow L}(E)$ (blue) for tip-to-sample (solid) and sample-to-tip (dashed) transport.
(b) The corresponding tunneling spin polarization $P_s$ for tip-to-sample ($P_s < 0$, red) and sample-to-tip ($P_s > 0$, blue). Reversing the bias polarity reverses the sign of $P_s$.
}
\label{fig:direction}
\end{figure}

\section{Bias-polarity-induced reversal of spin selectivity}
\label{sec:direction}

As discussed above, Eq.~\eqref{eq:spinPs-reversal} captures a unique property of topological-insulator tips that is absent in conventional ferromagnetic tips: the sign reversal of spin selectivity with bias polarity. To confirm this distinctive prediction, we now examine the bias dependence of the spin polarization $P_s$. As shown in Fig.~\ref{fig:direction}, reversing the STM bias polarity reverses the sign of $P_s$. Owing to the geometric asymmetry and spin-relaxing dephasing, the shorter lower edge remains the more transmissive channel for both transport directions. However, because of the spin-momentum locking of the QSH edge states, reversing the propagation direction reverses the spin orientation carried by the electrons along each edge [Fig.~\ref{fig:direction}(a)]. Consequently, the geometrically favored lower-edge path carries opposite spins under opposite biases, leading to a sign reversal of $P_s$, although its magnitude may differ due to the self-consistent probe potentials [Fig.~\ref{fig:direction}(b)]. This bias-driven sign reversal is consistent with the spin-resolved STM experiments using topological insulator tips~\cite{Aishwarya2022,Wang2025HelicalTip,Yang2026AltermagnetTip}.

Crucially, our results show that this spin selectivity is predominantly carried by the helical edge states. As shown in Fig.~\ref{fig:direction}(b), $|P_s|$ remains robust and finite as long as the electron energy $E$ lies within the bulk gap ($|E| < 0.5$), where the gapless helical edge states serve as the exclusive transport channels. Once $E$ lies outside the bulk gap, topologically trivial bulk modes begin to participate in the tunneling process. Because these bulk modes are largely immune to the edge path asymmetry, they carry no net spin selectivity, thereby diluting the total spin polarization and causing $|P_s|$ to drop rapidly towards zero. This sharp contrast between the edge-dominated and bulk-participating regimes provides compelling evidence that the observed spin polarization originates fundamentally from the path-asymmetric helical edge transport.

This strong dependence of $P_s$ on the topological boundary states is consistent with the SmB$_6$ tip experiment~\cite{Aishwarya2022}, where the spin contrast vanishes concomitantly with the suppression of the topological surface states above $\sim 10$ K, establishing topological electronic states as the essential prerequisite for the spin-selective tunneling. Furthermore, our transport framework could reproduce the magnetic stripe contrast and its bias-induced reversal observed in the experiment [see the SM~\cite{SM}].

\section{Geometry-guided tip design}
\label{sec:design}

\begin{figure}[t]
\centering
\includegraphics[width=0.48\textwidth]{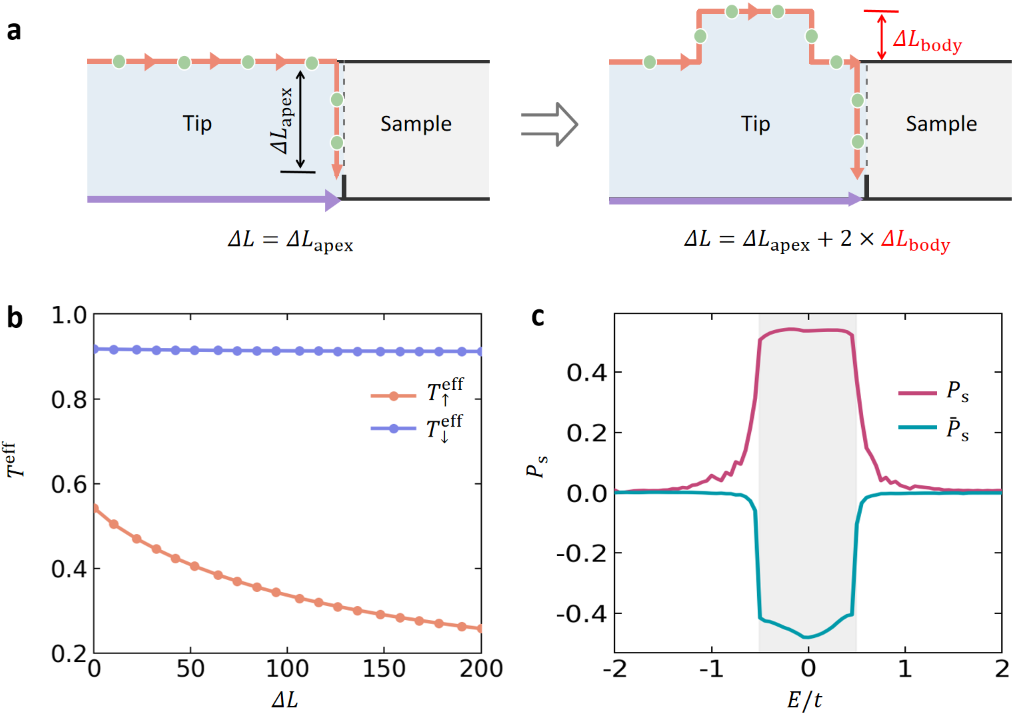}
\caption{Geometry-controlled enhancement of tunneling spin polarization.
(a) Significant enlargement of the edge-path difference $\Delta L$: arising solely from the tip apex (left) versus further extended by a geometric modification of the tip body (right).
(b) Spin-up transmission decreases with increasing $\Delta L$ toward an Ohmic regime, while spin-down transmission along the clean lower edge remains quantized.
(c) Tunneling spin polarization for positive ($P_s$) and negative ($\bar{P}_s$) bias polarities at $\Delta L = 100$, reaching $\sim 50\%$ in magnitude.
}
\label{fig:design}
\end{figure}

The contact asymmetry that drives the spin selectivity need not be limited to the atomic-scale tunneling apex: the tip body itself provides a practical knob for controlling helical edge transport. Because the spin-flip backscattering accumulated by a helical mode grows with its propagation distance, we can establish a simple design principle to selectively attenuate one helical edge channel by extending its length, while leaving the other channel intact. As illustrated in Fig.~\ref{fig:design}, we geometrically extend the upper edge of the tip body, such that the total edge-path difference $\Delta L$ now combines this macroscopic contribution with the atomic-scale apex asymmetry. Spin-relaxing dephasing centers act along the extended upper edge, whereas the shorter lower edge remains clean. With increasing $\Delta L$, the spin-up transmission along the extended edge is progressively suppressed. In stark contrast, the spin-down transmission along the clean lower edge remains close to its quantized value, characteristic of ballistic helical edge transport. The resulting imbalance strongly amplifies the spin polarization, which exceeds $50\%$ in our calculations.

This strategy bypasses the need for atomically precise engineering of the tunneling apex. Since some degree of spin relaxation is inevitable during edge propagation in realistic materials, the engineered path imbalance translates directly into strong spin selectivity, without microscopic control over the scatterer distribution. Macroscopic tip geometry thus emerges as an accessible and robust route to optimizing the spin sensitivity of topological STM probes.

\section{Effect of spin-orbit coupling}
\label{sec:rashba}

In the preceding sections where spin-orbit coupling (SOC) is absent, the tunneling spin polarization $P_s(E)$ is strictly defined by Eq.~\eqref{eq:spin_polarization} within the $\sigma_z$-conserving basis. However, realistic topological insulators may also host SOC, which breaks this conservation. To quantify this effect, we add a Rashba SOC term to the QSH tip,
\begin{equation}
H_R = i\frac{\alpha_R}{2}\sum_{\bm r}
\Big(c^\dagger_{\bm r+\hat{x}}\sigma_y\tau_0 c_{\bm r}-c^\dagger_{\bm r+\hat{y}}\sigma_x\tau_0 c_{\bm r} \Big) +\mathrm{H.c.},
\label{eq:rashba}
\end{equation}
where $\alpha_R$ is the SOC strength. Here, we only consider spin-relaxing dephasing. With finite $\alpha_R$, the composite terminal-spin indices used previously are no longer applicable. Instead, Eqs.~\eqref{eq:green_trans} and \eqref{eq:current-itoj} describe the total tunneling from terminal $i$ to terminal $j$, where $i=\{L, R, 1, 2, \cdots, N\}$ now denotes only the physical lead or probe index, and the spin degree of freedom must be explicitly traced out within each terminal. Consequently, the total current collected by the right lead is $I_R(E) = (e^2/h) T^{\rm eff}_{R \leftarrow L}(E) V$ with $T^{\rm eff}_{R \leftarrow L}(E) = \text{Tr}[\Gamma_R {\cal T}_R(E)]$, where we define the effective transmission matrix as
\begin{align}
{\cal T}_R(E) = G^r(E) \left[\Gamma_L + \sum_p \Gamma_{p} \frac{V_p}{V} \right] G^a(E).
\end{align}
The probe voltages $V_p$ are obtained by solving Eq.~\eqref{eq:current-itoj}. To further extract the spin-resolved contributions from $T^{\rm eff}_{R \leftarrow L}(E)$, we define a reduced matrix by taking a partial trace over the spatial and orbital degrees of freedom within the central region:
\begin{align}  \label{eq:spin_matrix}
\mathcal{S}(E) = \text{Tr}' \left[ \Gamma_R^{1/2} \mathcal{T}_R(E) \Gamma_R^{1/2} \right],
\end{align}
where $\text{Tr}'$ traces over the spatial and orbital degrees of freedom while retaining the spin degrees of freedom. Consequently, $\mathcal{S}(E)$ is a $2 \times 2$ Hermitian matrix in spin space that encodes the full spin-resolved transmission properties of the system~\cite{NikolicSouma2005}. Thus, the tunneling spin polarization along the $\sigma_i$ spin direction ($i=x,y,z$) is naturally defined as the normalized trace:
\begin{equation}\label{eq:PiSOC}
P_i(E) = \text{Tr}[\sigma_i \mathcal{S}(E)] / \text{Tr}[\mathcal{S}(E)].
\end{equation}
Here, the denominator $\mathrm{Tr}[\mathcal{S}(E)] = T^{\mathrm{eff}}_{R \leftarrow L}(E)$ represents the total transmission, while the numerator evaluates the spin-resolved transmission difference. In the SOC-free limit, the \(\sigma_z\) conservation ensures that \(\mathcal{S}(E)\) is strictly diagonal in spin space (i.e., \(\mathcal{S}_{\uparrow\downarrow} = \mathcal{S}_{\downarrow\uparrow} = 0\)). Consequently, \(P_z(E)\) exactly reduces to the original definition \(P_s(E) = (\mathcal{S}_{\uparrow\uparrow} - \mathcal{S}_{\downarrow\downarrow})/(\mathcal{S}_{\uparrow\uparrow} + \mathcal{S}_{\downarrow\downarrow})\) given in Eq.~\eqref{eq:spin_polarization}.

\begin{figure}[t]
\centering
\includegraphics[width=0.48\textwidth]{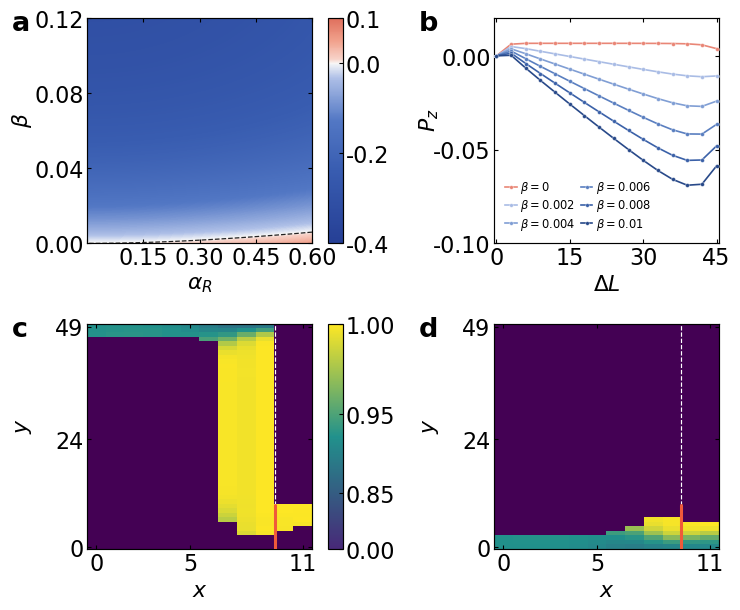}
\caption{Effect of Rashba SOC on tunneling spin polarization. 
(a) Spin polarization $P_z$ in the $\alpha_R$-$\beta$ plane at $E=0.01t$; the dashed contour marks $P_z=0$. Spin dephasing dictates the dominant polarization, whereas Rashba SOC acts only as a weak competing mechanism. 
(b) $P_z$ as a function of $\Delta L$ at $\alpha_R=0.2$ for various $\beta$.
(c,d) Real-space maps of the local spin polarization $P_z(\mathbf{r})$ for the (c) top and (d) bottom edge states at $\beta=0$ [corresponding to the orange curve in (b)], isolating the pure SOC effect.
}
\label{fig:rashba}
\end{figure}

We present the tunneling spin polarization $P_z(E)$ in Figs.~\ref{fig:rashba}(a) and (b) to discuss the role of SOC within our specific model. Figure~\ref{fig:rashba}(a) maps $P_z$ in the $\alpha_R$-$\beta$ plane at $E=0.01$. In the absence of SOC, the asymmetric tunneling-contact geometry ($\Delta L\neq 0$) and spin-relaxing dephasing ($\beta \neq 0$) yield a negative spin polarization [Fig.~\ref{fig:controls}]. A finite Rashba SOC introduces a competing spin-mixing effect that drives $P_z$ towards positive values; however, the dephasing-induced polarization remains dominant across the entire parameter range considered. Moreover, we investigate how $\Delta L$ modulates this competition in Fig.~\ref{fig:rashba}(b) for a fixed $\alpha_R=0.2$. As the contact length decreases (i.e., as $\Delta L$ increases), spin-relaxing dephasing ($\beta \neq 0$) rapidly induces a net negative polarization, outweighing the weak positive polarization produced by pure SOC ($\beta = 0$). Crucially, both contributions vanish completely in the symmetric contact limit, confirming that geometric asymmetry is the essential prerequisite for finite spin polarization.

The positive polarization generated by pure SOC ($\beta=0$) can be understood from the local spin-polarization maps in Figs.~\ref{fig:rashba}(c) and (d), with calculation details provided in the SM~\cite{SM}. Because the horizontal (upper/lower) and vertical (right) edges have different crystalline orientations, the Rashba term induces an asymmetric spin rotation along the boundaries. For our QSH model, when reaching the contact, the electrons traversing the longer upper-edge path exhibit a stronger out-of-plane spin component compared to those on the shorter lower edge. The right-moving electrons along the bottom edge therefore lose more of their out-of-plane spin polarization, resulting in $I_\uparrow > I_\downarrow$ and thereby yielding a net positive $P_z$. While we focused on Rashba SOC in the BHZ model, exploring other SOC types or topological insulator models is left for future work.

\section{Conclusion and discussions}

In summary, we have developed a spin-resolved transport theory of scanning tunneling microscopy with topological-insulator tips. We bridge the Bardeen tunneling picture and the mesoscopic Landauer-B\"uttiker framework, establishing a unified Bardeen-Landauer-B\"uttiker formalism for topological quantum probes. Our central finding is a cooperative mechanism for generating a net spin-polarized tunneling current: it requires the coexistence of an asymmetric tunneling-contact geometry (e.g., the tip apex) and spin-relaxing dephasing (e.g., magnetic impurities or spin-orbit scattering). While the geometric asymmetry creates helical edge paths of unequal lengths, the spin dephasing provides the necessary backscattering to break the balance between the counter-propagating channels. This joint effect produces unequal spin-resolved transmissions and a finite net polarization. Our predictions are qualitatively consistent with recent spin-selective STM experiments using SmB$_6$ and Bi$_4$Br$_4$ tips~\cite{Aishwarya2022,Wang2025HelicalTip,Yang2026AltermagnetTip}. 

The resulting spin selectivity exhibits distinctive features that separate topological tips from conventional ferromagnetic tips. The polarization is strictly carried by the helical edge states within the bulk gap and naturally reverses its sign when the STM bias polarity is switched, directly reflecting the spin-momentum locking. Furthermore, this mechanism is robust against the spin-mixing effect induced by Rashba SOC. Notably, the polarization might be enhanced (exceeding $50\%$) by extending the macroscopic tip body, providing a practical route for geometry-guided probe design that bypasses the stringent requirement of atomic-scale apex engineering. 

Our transport-based mechanism provides a route to bias-tunable spin selectivity without equilibrium tip magnetization, which is distinct from both electric-field-induced magnetoelectric responses~\cite{Banerjee2025} and interaction-induced surface magnetism~\cite{Dzero2026OddParity,Dzero2026SpinPolarized}. Looking forward, generalizing our work to other topological phases will offer new design principles for tunable spin-selective probes. In particular, layered topological materials such as HfTe$_5$ and ZrTe$_5$, which host helical boundary modes, represent promising candidates for realizing experimentally accessible topological probes~\cite{Weng2014QSH,Liu2019HfTe5,Wu2016ZrTe5}.


\bibliography{references_updated}

\end{document}